\documentclass[prb,aps,twocolumn]{revtex4}
\usepackage{graphicx,color,latexsym}
\usepackage{dcolumn}
\usepackage{amsmath,amssymb,epsf,bm}
\begin{document}
\title{Generation of electric current and electromotive force  by an  antiferromagnetic domain wall}
\author{A.~G. Mal'shukov}
\affiliation{Institute of Spectroscopy, Russian Academy of Sciences, Troitsk, Moscow, 108840, Russia}

\begin{abstract}
Dynamic magnetic textures may transfer the angular moment from the varying in time antiferromagnetic order to spins of conduction electrons. Due to the spin orbit coupling (SOC) these spin excitations can induce the electric current of conduction electrons. We calculated the electric current and the electromotive force (EMF) which are produced by a domain wall (DW) moving parallel to the  magnetically compensated interface between an antiferromagnetic insulator (AFMI) and a two-dimensional spin orbit coupled metal. Spins of conduction electrons interact with localized spins of a collinear AFMI through the interface exchange interaction.  The Keldysh formalism of nonequilibrium Green functions was applied for the analysis of this system. It is shown that a Bloch DW generates the current perpendicular to the DW motion direction. At the same time a N\'{e}el DW creates the electric potential which builds up across the wall. The total charge which is  pumped by a Bloch DW can be expressed in terms of a topologically invariant charge quantum. The latter does not depend on variations of DW's velocity and shape. These effects increase dramatically when the Fermi energy approaches the van Hove singularity of the Fermi surface. The obtained results are important for the electrical detection and control of dynamic magnetic textures in antiferromagnets.
\end{abstract}
\maketitle

\section{Introduction}

Antiferromagnets (AFM) have drawn growing interest recently due to their potential use for various spintronic applications. Compared with ferromagnets, in AFM spin waves and topological spin textures exhibit much faster dynamics. These materials demonstrate remarkable capabilities to transmit spin polarization over large distances. Moreover, insulating and metallic AFM's can efficiently transfer the spin polarization through interfaces with paramagnetic metals. The experimental and theoretical progress made in AFM spintronics has been presented  in a number of reviews.\cite{Baltz,Gomonay,Yan,Brataas}  Insulating materials are of particular interest within the AFM family \cite{Brataas}, because energy losses which are associated with spin waves and other dynamical spin textures are weak in these materials. The low dissipative spin transport is one of the most important advantages of magnetic insulators. For example, spin waves can propagate there over several microns \cite{Cornelissen,Lebrun} New avenues in AFMI spintronics are opened due to fabrication of heterostructures which combine magnetic insulators and paramagnetic metals. In such nanostructures efficient ways of manipulating and detecting spin transport become available, based on experimental and theoretical progress in understanding of mechanisms of the angular moment transfer between magnetic insulators and metals. \cite{Cheng,Vaidya,Li,Wang} These studies were mostly aimed at magnons, as carriers of the spin polarization. On the other hand, topological spin textures, such as domain walls and skyrmions, also can efficiently transfer the spin polarization. They are of great importance for various spintronic applications due to their high mobility and non-volatility. As for DW, a considerable progress has been made in understanding and experimental implementation of various DW propulsion mechanisms which are important for DW manipulation in AFMI, where DW velocity reaches rather high values, which are much larger than in ferromagnetic materials  \cite{Caretta,Siddiqui,Avci,Zhou,Velez,Gomonay2,Kim} However, mechanisms which enable the electric control and detection of DW's dynamics  require further studies.

In this work the electric detection of a moving DW will be considered for a heterostructure consisting of an AFMI layer which makes a magnetically compensated  contact with a 2D metal film. In such a hybrid system, due to proximity with AFMI, the 2D gas of electrons behaves as an AFM metal. The effect of DW dynamics on conduction electrons has earlier been considered for ferromagnets \cite{Stern,Barnes,Duine,Tserkovnyak2}. There, DW motion could induce the electric current and EMF due to a difference between conductivities of electrons with opposite spins. Such a mechanism cannot operate in a compensated AFM. In this case the conversion of the time dependent AFM order into electricity can be provided by SOC.  As known \cite{Cheng,Saidaoui,Swaving,Takei,Nunez,Ohnuma},  the flux of magnons, which is incident on the interface between AFMI and a heavy metal, can generate there the electric current. The latter is induced by the inverse spin-Hall effect which is caused by the spin current from the interface into the bulk of the metal. Such an effect  also takes place when the angular moment from various dynamic spin textures is transferred through the interface. However, this mechanism can not operate, if the AFMI is contacted to a 2D metal film. Therefore, in this situation we shall consider the generation of the electric current and EMF  which directly involves the interface exchange interaction $J$ between 2D electrons and localized spins of the AFMI. The conversion of the angular moment, which is carried by a DW,  into electric current and electric potential is provided by the Rashba \cite{Rashba} SOC. The latter may be very strong at the interface where  the inversion symmetry is violated.

The electric current will be calculated  within the Keldysh \cite{Keldysh} formalism of nonequilibrium Green's functions. This problem will be considered within a simple tight binding model where  2D metal and AFMI lattices form a commensurate contact. It will be assumed that during DW's motion the N\'{e}el order varies smoothly in time and space, in comparison with the conduction electron's dynamics near the Fermi level. Therefore, these slow variations will be treated perturbatively. It was found that the time dependent variation of the N\'{e}el order, which is caused by a moving DW, produces a qualitatively different and much stronger effect than spatial variations. While the latter adiabatically modify  wave functions and energies of electrons, the former mechanism produces excited electrons near the Fermi level. Therefore, it modifies the electron's distribution function, rather than their spectrum. Previously, a similar approach was employed for studying the spin accumulation which can be caused by a moving DW.\cite{Malshukov1} There, however, the exchange interaction of conduction electrons with localized spin was assumed to be small in comparison with the distance of the electron's Fermi energy from the van Hove singularity. Now, these energies are of the same order. Their closeness leads to a strong enhancement of the effects which are produced by the DW.

The article is organized in the following way. In Sec.II a general formalism is presented. Sec. III is devoted to calculations of the electrical current and EMF which are induced by N\'{e}el  and Bloch DW's, respectively. The results are discussed in Sec. IV. Three sections are added to the Appendix, in order to clarify some details of calculations.

\begin{figure}[tp]
\includegraphics[width=7 cm]{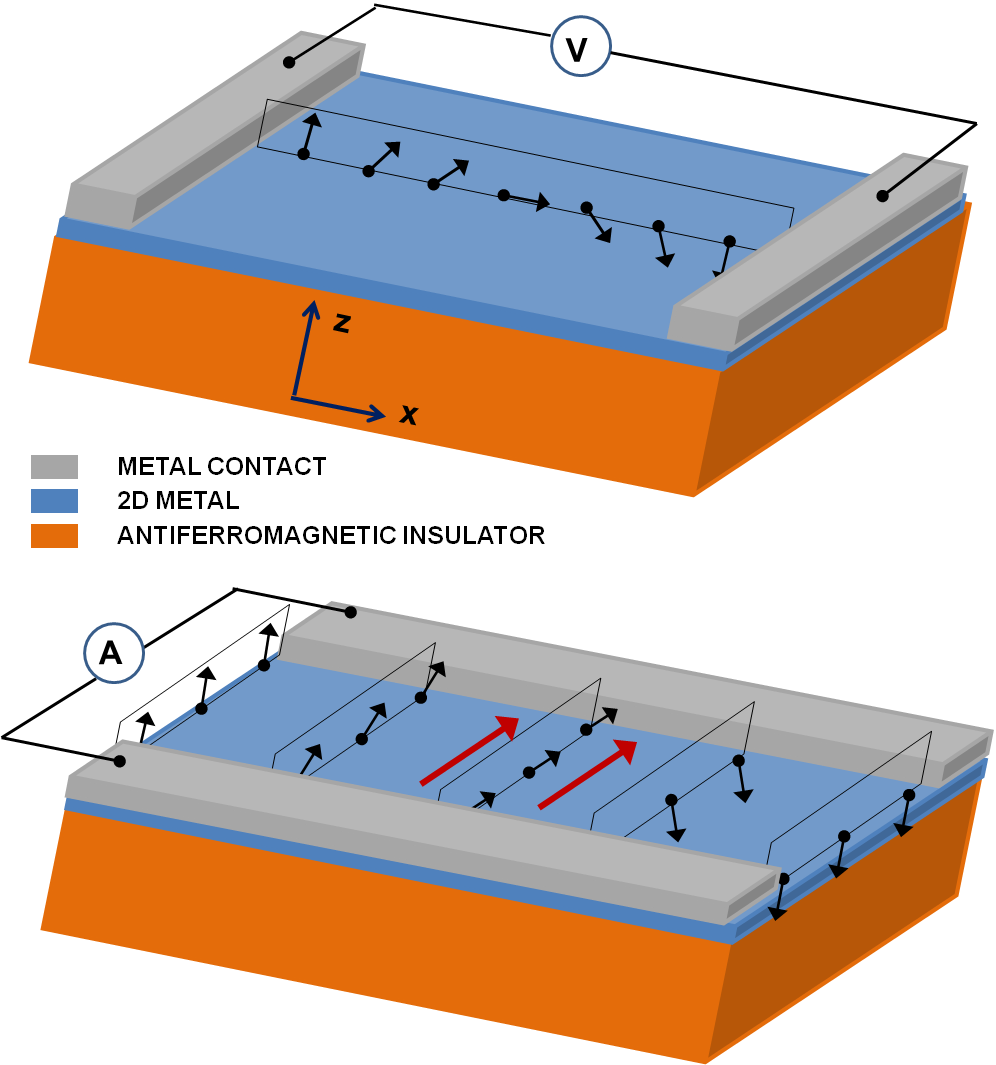}
\caption{(Color online) The electric current and electromotive force are induced in the 2D metal film by a domain wall which moves  in the antiferromagnetic insulator in the x-direction. Top: the electric potential builds up across the N\'{e}el domain wall (the N\'{e}el order is shown by arrows). The corresponding voltage V can be measured by a voltmeter in the open circuit. Bottom: the moving Bloch domain wall generates  the electric current close to it (shown by thick arrows), which is directed perpendicular to the DW motion direction. This current can be measured by an amperemeter in the closed circuit} \label{fig1}
\end{figure}

\section{General formalism}

In this section the electric current in 2D electron gas will be expressed  in terms of the nonequilibrium Keldysh Green's function which, in turn, can be obtained as an expansion over temporal and spatial gradients of the AFM N\'{e}el order. By assuming that the 2D lattice of the normal metal is commensurate with the lattice of localized spins on the AFMI interface and that metal atoms make an on-top contact with atoms of the AFMI, the exchange interaction can be written in the form
\begin{equation}\label{Hint}
H_{\mathrm{ex}}(t)=J\sum_i \psi^{\dag}(\mathbf{r}_i)\bm{\sigma}\cdot\mathbf{S}(\mathbf{r}_{i},t)\psi(\mathbf{r}_i) \,.
\end{equation}
where $\psi^{\dag}(\mathbf{r}_i)=(\psi^{\dag}_{\uparrow}(\mathbf{r}_i),\psi^{\dag}_{\downarrow}(\mathbf{r}_i))$ is the two-component creation operator of an electron whose spin projections are $\uparrow$, or $\downarrow$ and $\psi(\mathbf{r}_i)$ is the corresponding destruction operator.  The vector $\mathbf{S}(\mathbf{r}_{i},t)$ represents a  spin which is localized on the lattice site $\mathbf{r}_i$, while $\bm{\sigma}=(\sigma_x,\sigma_y,\sigma_z)$ is the vector of Pauli matrices. Localized  spins are treated as classical variables satisfying the constraint $|\mathbf{S}(\mathbf{r}_{i},t)|=S$. In many practical situations $\mathbf{S}(\mathbf{r}_{i},t)$ varies slowly  within each of two AFM sublattices A and B. Therefore,  one may introduce two vector fields $\mathbf{m}_A(\mathbf{r},t)$ and $\mathbf{m}_B(\mathbf{r},t)$, which are defined on sublattices A and B, respectively, where $\mathbf{m}_{A(B)}(\mathbf{r}_{i},t)=\mathbf{S}(\mathbf{r}_{iA(B)})/S$. The N\'{e}el order is given by the unit vector field  $\mathbf{n}(\mathbf{r},t)=(\mathbf{m}_A(\mathbf{r},t) - \mathbf{m}_B(\mathbf{r},t))/|\mathbf{m}_A(\mathbf{r},t) - \mathbf{m}_B(\mathbf{r},t)|$. Due to the strong exchange coupling of spins in different sublattices we have $\mathbf{m}_A(\mathbf{r},t) \simeq -\mathbf{m}_B(\mathbf{r},t)$. It is convenient to label sublattices by a new variable $\tau$, such that $\tau=1$ and $\tau=-1$ on sublattices A and B, respectively. Accordingly, by introducing the vector of Pauli matrices  $\bm{\tau}=(\tau_x,\tau_y,\tau_z)$ Eq.(\ref{Hint}) may be written in the form
\begin{equation}\label{Hint2}
H_{\mathrm{ex}}(t)=JS\sum_{i}\psi^{\dag}(\mathbf{r}_i)\mathbf{n}(\mathbf{r}_i,t)\cdot\bm{\sigma}\tau_z \psi(\mathbf{r}_i) \,,
\end{equation}
where electron creation operators are defined as $\psi^{\dag}(\mathbf{r}_i)=(\psi^{\dag}_{\uparrow,1}(\mathbf{r}_i),\psi^{\dag}_{\downarrow,1}(\mathbf{r}_i),\psi^{\dag}_{\uparrow,-1}(\mathbf{r}_i),\psi^{\dag}_{\downarrow,-1}(\mathbf{r}_i))$. The unperturbed Hamiltonian $H_{\mathrm{tb}}$  is represented by the tight binding model which takes account of near-neighbor electron's hopping and the Rashba  spin-orbit coupling. It has the form \cite{Zelezny}
\begin{eqnarray}\label{Htb}
&&H_{\mathrm{tb}}=-t\sum_{i\lambda\bm{\delta}_{\gamma}} \psi^{\dag}(\mathbf{r}_i)\tau_x \psi(\mathbf{r}_i+\lambda\bm{\delta}_{\gamma}) +\nonumber\\
&&\lambda_R\sum_{i\lambda} \lambda\psi^{\dag}(\mathbf{r}_i)\tau_x[\sigma_y\psi(\mathbf{r}_i+\lambda\bm{\delta}_x)- \sigma_x\psi(\mathbf{r}_i+\lambda\bm{\delta}_y)]\,,
\end{eqnarray}
where $\lambda=\pm1$, $\bm{\delta}_{\gamma}=a\mathbf{e}^{\gamma}$ with $\gamma=x,y$ denoting the orientation of the unit vectors $\mathbf{e}^{\gamma}$, $a$ is the lattice constant, $t$ is the hopping amplitude, and  $\lambda_R$ is the SOC constant. The total Hamiltonian can be written as
\begin{equation}\label{H}
H=\sum_{ij} \psi^{\dag}(\mathbf{r}_i)(\mathcal{H}_{\mathrm{tb}ij}+\mathcal{H}_{\mathrm{ex}ij}(t)) \psi(\mathbf{r}_j) \,,
\end{equation}
where $\mathcal{H}=\mathcal{H}_{\mathrm{tb}ij}+\mathcal{H}_{\mathrm{ex}}$ is the one-particle  Hamiltonian of electrons. At the same time, $\mathcal{H}_{\mathrm{tb}ij}$ can be written in the form $\mathcal{H}_{\mathrm{tb}ij}=\mathcal{H}_{0ij}+\mathcal{H}_{Rij}$, where the $\mathcal{H}_{0ij}$ and $\mathcal{H}_{Rij}$ correspond to the first (spin independent) and the second (spin dependent) terms of Eq.(\ref{Htb}), respectively. The SOC term $\mathcal{H}_{Rij}$ can be written in the form $\mathcal{H}_{Rij}=\mathbf{h}_{Rij}\bm{\sigma}\tau_x$.

From Eq.(\ref{Htb}) one can write the equation for retarded and advanced Green's functions as
\begin{equation}\label{G}
\sum_{l}\left(-i\delta_{il}\frac{\partial }{\partial t}-\mathcal{H}_{il}+\mu\delta_{il}\right)G_{lj}(t.t^{\prime})=\delta_{ij}\delta(t-t^{\prime}) \,,
\end{equation}
where $\mu$ is the chemical potential. It is convenient to modify Eq.(\ref{G}) by using the SU(2) unitary transformation which locally aligns the spin quantization axis of electrons with the N\'{e}el  vector. Let us consider a one-dimensional DW whose N\'{e}el vector  is given by the well known solution of the equation of motion for a  DW in an uniaxial AFM.\cite{Walker} This wall moves in the $x$-direction and its N\'{e}el  vector evolves within a plane which is perpendicular to the $xy$ plane and whose normal is determined by the unit vector $\bm{\nu}$. Then, the unitary transformation $U^{-1}\mathbf{n}(\mathbf{r},t)\bm{\sigma}U$, where $U=\exp\left(i\bm{\nu}\bm{\sigma}\theta(x,t)/2\right)$ and $\theta(x,t)$ is the polar angle of the N\'{e}el  vector $\mathbf{n}(x,t)$, results in $\mathbf{n}(x,t)\bm{\sigma}\rightarrow \sigma_z$. By applying this operation to Eq.(\ref{G}) we obtain  a transformed equation which contains gradients of  $\theta(x,t)$. For instance,
\begin{eqnarray}\label{H0prime}
&&e^{-i\bm{\nu}\bm{\sigma}\theta(x_i,t)/2}\mathcal{H}_{\mathrm{tb}ij}e^{i\bm{\nu}\bm{\sigma}\theta(x_j,t)/2}= \mathcal{H}^{\prime}_{\mathrm{tb}ij}-\nonumber\\
&&i\partial_x\theta(x_0,t)\left[v^x_{ij}\frac{\bm{\nu}\bm{\sigma}}{2}+\frac{\{\bm{\nu}\bm{\sigma},v^{xy}_{ij}\}}{4}\right]\,,
\end{eqnarray}
where  $x_0=(x_i+x_j)/2$, $v^x_{ij}=-ta\tau_x\lambda\delta_{\mathbf{r}_j,\mathbf{r}_i+\lambda\bm{\delta}_{x}}$ and $v^{xy}_{ij}=a\lambda_R\tau_x\sigma_y\delta_{\mathbf{r}_j,\mathbf{r}_i+\lambda\bm{\delta}_{x}}$. Here, $v^x_{ij}$ and $v^{xy}_{ij}$ are the velocity and spin-velocity operators, respectively. Since these operators depend on the coordinate difference $\mathbf{r}_j-\mathbf{r}_i$ they can be written in the momentum representation as $v^x=\mathrm{Tr}[\nabla_{k_x}\mathcal{H}_{\mathbf{k}\mathrm{tb}}]/2$ and $v^{xy}=\sigma_y\mathrm{Tr}[\sigma_y\nabla_{k_x}\mathcal{H}_{\mathbf{k}\mathrm{tb}}]/2$, where the Hamiltonian $\mathcal{H}_{\mathbf{k}\mathrm{tb}}$ can be obtained from Eq.(\ref{Htb}) as
\begin{equation}\label{Hktb}
\mathcal{H}_{\mathbf{k}\mathrm{tb}}=-2t(\cos ak_x+\cos ak_y)\tau_x+\tau_x\bm{\sigma}\mathbf{h_{R\mathbf{k}}}\,,
\end{equation}
with $h_{R\mathbf{k}}^x=-2\lambda_R\sin ak_y$ and $h_{R\mathbf{k}}^y=2\lambda_R\sin ak_x$.
The  Hamiltonian $ \mathcal{H}^{\prime}_{\mathrm{tb}ij}= \mathcal{H}^{\prime}_{0ij}+ \mathcal{H}^{\prime}_{\mathrm{R}ij}$ in Eq.(\ref{H0prime}) does not contain gradient terms. It is obtained by a local $SU(2)$ rotation of spin operators in Eq.(\ref{Htb}).  As a result, only the spin dependent Rashba Hamiltonian $\mathcal{H}_{\mathrm{R}ij}$ in Eq.(\ref{Htb}) is modified. It is convenient to project spin operators $\sigma_x$ and $\sigma_y$ onto two orthogonal directions $\bm{\nu}$ and $\overline{\bm{\nu}}=(\mathbf{e}_z\times\bm{\nu})$, so that $\bm{\sigma}=(\bm{\nu}\bm{\sigma})\bm{\nu}+(\overline{\bm{\nu}}\bm{\sigma})\overline{\bm{\nu}}$. The first operator in this equation stays invariant under the $SU(2)$ transformation, while the second one becomes  time and space dependent. The so transformed operators will be denoted as $\bm{\sigma}^{\nu}$ and  $\bm{\sigma}^{\overline{\nu}}$, respectively, where $\bm{\sigma}^{\overline{\nu}}=U^+(\mathbf{r},t)(\overline{\bm{\nu}}\bm{\sigma})\overline{\bm{\nu}}U(\mathbf{r},t)$.

After the unitary transformation Eq.(\ref{G}) takes the form
\begin{eqnarray}\label{Gtilde}
&&\sum_{l}\left(-i\delta_{il}\frac{\partial }{\partial t}-\mathcal{H}^{\prime}_{0il}-\mathbf{h}^{\nu}_{Rij}\bm{\sigma}\tau_x-\delta_{il}JS\tau_z\sigma_z+\mu\delta_{il}-\right.\nonumber\\
&&\left.\mathbf{h}^{\overline{\bm{\nu}}}_{Rij}(x_0,t)\bm{\sigma}\tau_x +\delta_{il}\mathcal{A}_0+\mathcal{A}_xv^x_{il}+ \frac{1}{2}\{\mathcal{A}_x,v^{xy}_{il}\}\right)\times\nonumber \\
&&\tilde{G}_{lj}(t.t^{\prime})=\delta_{ij}\delta(t-t^{\prime})
\end{eqnarray}
where $\tilde{G}=U^{+}GU$, $\mathcal{A}_0=\bm{\nu}\bm{\sigma}\partial_t\theta(x,t)/2$, $\mathcal{A}_x=\partial_x\theta(x,t)\bm{\nu}\bm{\sigma}/2$, $\mathbf{h}^{\overline{\bm{\nu}}}_{Rij}(x_0,t)=(\overline{\bm{\nu}}\mathbf{h}_{Rij})(\overline{\bm{\nu}}\sin\theta(x_0,t)+\mathbf{e}_z\cos\theta(x_0,t))$ and $\mathbf{h}^{\nu}_{Rij}=(\mathbf{h}^{\nu}_{Rij}\bm{\nu})\bm{\nu}$. This equation allows to calculate the retarded and advanced Green's functions by using the perturbation theory with respect to small gauge fields and the weak Rashba coupling. The unperturbed functions $\tilde{G}_0^{r(a)}$ are determined by space-time independent terms of the Hamiltonian, which are collected in the top line of Eq.(\ref{Gtilde}). In the momentum representation the corresponding  Hamiltonian is given by
\begin{equation}\label{Hk0}
\mathcal{F}_{\mathbf{k}}=\tau_x\epsilon_{\mathbf{k}}+\tau_x(\bm{\nu}\bm{\sigma})(\bm{\nu}\mathbf{h}_{R\mathbf{k}})+JS\tau_z\sigma_z-\mu\,,
\end{equation}
where $\epsilon_{\mathbf{k}}=-2t(\cos ak_x+\cos ak_y)$. The unperturbed Green functions $\tilde{G}_{\mathbf{k}0}^{r(a)}$ can be immediately found as $(\omega-\mathcal{F}_{\mathbf{k}})^{-1}$,  at arbitrary strength of SOC in Eq.(\ref{Hk0}). We, however, present it in the form of the linear expansion with respect to SOC, because there is the space-time dependent part of this coupling which is presented in Eq.(\ref{Gtilde}), but does not enter in Eq.(\ref{Hk0}). It may be taken into account perturbatively, as shown in Appendix A. Therefore, it is reasonable to treat these two parts of SOC similarly. From Eq.(\ref{Hk0}) $\tilde{G}_0^{r(a)}$ can be expressed in the form
\begin{equation}\label{G0}
\tilde{G}_{\mathbf{k}0}^{r(a)}=g_{\mathbf{k}}^{r(a)}(\omega)+g_{\mathbf{k}}^{r(a)}(\omega)\tau_x(\bm{\nu}\bm{\sigma})(\bm{\nu}\mathbf{h}_{R\mathbf{k}})g_{\mathbf{k}}^{r(a)}(\omega)\,,
\end{equation}
where
\begin{equation}\label{G0A}
g_{\mathbf{k}}^{r(a)}(\omega)=\frac{1}{2}\sum_{\beta=\pm1}\frac{(1+\beta \hat{P}_{\mathbf{k}})}{\omega+\mu-\beta E_{\mathbf{k}}\pm i\Gamma}.
\end{equation}
In this equation $E_{\mathbf{k}}=\sqrt{\epsilon^2_{\mathbf{k}}+J^2S^2}$ and $\hat{P}_{\mathbf{k}}=(\epsilon_{\mathbf{k}}\tau_x+JS\tau_z\sigma_z)/E_{\mathbf{k}}$. The parameter $\Gamma$ is infinitesimally small in a clean 2D system. However, we will take into account some elastic impurity scattering of electrons in the Born approximation, so that $\Gamma$ becomes finite (in more detail it is discussed in Appendix C). The poles of $g_{\mathbf{k}}^{r(a)}(\omega)$ determine spin degenerate band energies in an antiferromagnet with the uniform N\'{e}el order and Rashba SOC. These energies have the gap $\Delta=JS$ in the middle of the band. All energies and the chemical potential are counted from this point.

Eqs.(\ref{Gtilde}) and (\ref{G0}) allow us to calculate retarded and advanced Green functions of the electronic system.  In general, these functions contain all information which is sufficient for studying effects produced by static spin textures, like in Ref.[\onlinecite{Swaving}]. However, for dynamic spin textures it is necessary to know the  electronic distribution function, or more generally, the density matrix. This matrix can be represented by the Keldysh  function $G^K$, which is given by $G^K=G^<+G^>$ \cite{Rammer}. For electrons which interact with a dynamic AFM texture $G^K$ may be expressed in terms of $G^r$, $G^a$ and the function $f(\omega)=\tanh(\omega/k_BT)=1-2n(\omega)$, where $n(\omega)$ is the  equilibrium Fermi distribution of electrons.  It is written as \cite{Kopnin} $G^K=G^{K1}+G^{K2}$, where
\begin{eqnarray}\label{GK1}
&&G^{K1}_{\mathbf{k,k-Q}}(\omega,\omega-\Omega)=G^r_{\mathbf{k,k-Q}}(\omega,\omega-\Omega)f(\omega-\Omega)-\nonumber\\
&&f(\omega)G^a_{\mathbf{k,k-Q}}(\omega,\omega-\Omega)\,,
\end{eqnarray}
and $G^{K2}$ is given by
\begin{eqnarray}\label{GK2}
&&G^{K2}_{\mathbf{k,k-Q}}(\omega,\omega-\Omega)=\sum_{\mathbf{Q}^{\prime},\mathbf{Q}^{\prime\prime}}\int \frac{d\Omega^{\prime} d\omega^{\prime}}{4\pi^2}G^r_{\mathbf{k,k}-\mathbf{Q}^{\prime}}(\omega,\omega^{\prime})\times\nonumber\\
&&V_{\mathbf{Q}^{\prime\prime}}(\omega^{\prime},\omega^{\prime}-\Omega^{\prime})G^a_{\mathbf{k}-\mathbf{Q}^{\prime}-
\mathbf{Q}^{\prime\prime},\mathbf{k-Q}}(\omega^{\prime}-\Omega^{\prime},\omega-\Omega)\,,
\end{eqnarray}
where
\begin{equation}\label{V}
V_{\mathbf{Q}}(\omega,\omega-\Omega)=
SJ\tau_z\bm{\sigma}\mathbf{n}_{\mathbf{Q}}(\Omega)(f(\omega)-f(\omega-\Omega))
\end{equation}
and $\mathbf{n}_{\mathbf{Q}}(\Omega)$ denotes the Fourier transform of the N\'{e}el order.
Note, that all Green's functions in these equations, namely, $G^r,G^a$ and $G^K$ depend on a DW spin texture which varies in time and space. Therefore, these functions are inhomogeneous, so that their Fourier transforms in Eqs.(\ref{GK1}-\ref{GK2}) depend on two wave vector arguments and  two frequency arguments. Accordingly, $\mathbf{Q}$ and $\Omega$ in these equations are associated with the wave vector and the frequency of the N\'{e}el order, respectively.

Eqs.(\ref{GK1}) and (\ref{GK2}), together with  Eqs.(\ref{Gtilde}) and (\ref{G0}) allow us to calculate nonequilibrium effects caused by a dynamic magnetic structure. Note, however, that Eq.(\ref{GK2}) is valid only at sufficiently weak disorder. The problem is that at the stronger disorder one should take into account the diffusion of particles and spins. Formally, this diffusion is expressed in terms of multiple scattering processes which lead to the renormalization of current vertices by series of ladder diagrams. Spin and particle diffusion processes are coupled to each other through the  SOC  \cite{Malshukov}. At weak disorder, when the scattering rate is much less than the SOC, the electron spin precession, which is caused by the Rashba field $h_R$, is fast, so that spins completely lose their orientation after a single scattering event. Therefore, the spin diffusion does not take place. At the same time, the particle diffusion can be ignored for current vertices at the small wave-number $\mathbf{Q}$ in Eq.(\ref{GK2}).\cite{agd}

\section{Electric current and EMF}

The electric current density can be expressed in terms of the Keldysh function as
\begin{equation}\label{j}
\mathbf{j}(\mathbf{r},t)=-\frac{ie}{2}\sum_{\mathbf{k,p}}\mathrm{Tr}[\mathbf{j}_{\mathbf{k}}G^K_{\mathbf{k},\mathbf{p}}(t,t)]e^{i\mathbf{(k-p)r}} \,,
\end{equation}
where $\mathbf{j}_{\mathbf{k}}$ is given by $\nabla_{\mathbf{k}}\mathcal{H}_{\mathbf{k}\mathrm{tb}}$, with $\mathcal{H}_{\mathbf{k}\mathrm{tb}}$ determined by Eq.(\ref{Hktb}). This current will be calculated for N\'{e}el and Bloch DW's. In these cases the unit vector $\bm{\nu}$ in  Eq.(\ref{H0prime}) is parallel to $y$ and $x$ axes, respectively. There is a qualitative difference between effects produced by these two types of  DW's. As will be shown, in the latter case the electric current is directed in the $y$-direction which is perpendicular to the motion of the DW. In the considered set up the current is homogeneous in this direction and may be conveniently collected by side terminals, as shown in Fig.1 (bottom). At the same time, a N\'{e}el DW induces  the current which is parallel to $\mathbf{e}^x$ and is inhomogeneous along this direction, so that near the DW the current conservation law is violated. In this case, however, the electric potential step is created inside the DW. This potential induces a countercurrent. As a result, in an open circuit the total current density vanishes. Instead, the potential difference is created between terminals, which are placed  in the front and rear of the DW, as shown in Fig.1 (top). Let us consider first the Bloch DW.

\subsection{Bloch DW}

The total current $J_y$ in the $y$-direction is obtained by integration of Eq.(\ref{j}) over $x$. Then, since the considered system is homogeneous along $y$, in so integrated  Eq.(\ref{j}) one can set $\mathbf{p=k}$. Moreover,  $J_y$ does not depend on time, if the DW  moves with a constant velocity. Therefore, in Eq.(\ref{GK1}) and Eq.(\ref{GK2}) $\Omega\rightarrow 0$, so that the total current, which is averaged over the time interval $T$ required for DW to pass through the sample, can be written as
\begin{equation}\label{Jy}
J_y=-\frac{ie}{2T}\sum_{\mathbf{k}}\int \frac{d\omega}{2\pi} \mathrm{Tr}[j^y_{\mathbf{k}}(G^{K1}_{\mathbf{k},\mathbf{k}}(\omega,\omega)+G^{K2}_{\mathbf{k},\mathbf{k}}(\omega,\omega))] \,,
\end{equation}
where $j^y_{\mathbf{k}}=\tau_x\partial_{k_y}(\epsilon_{\mathbf{k}}+\bm{\sigma}_x h^x_{R\mathbf{k}})$ First, let us consider  $J_y$ which stems from the first term in Eq.(\ref{Jy}) and denote the corresponding  current as $J_y^{(1)}$. The retarded and advanced functions in Eq.(\ref{GK1}) may be expressed as $U\tilde{G}^{r(a)}U^+$. For a Bloch DW the matrix $U$ is given by $U=\exp(\pm i\sigma_x\theta(x,t)/2)$. The sign of the exponent depends on a direction of the vector $\bm{\nu}$ in the unitary transformation $U=\exp(i\bm{\nu}\bm{\sigma}\theta/2)$, where $\bm{\nu}$ is parallel to the x-axis for a Bloch DW. Let us choose, for simplicity, the "+" sign, that corresponds to the positive $x$-direction of this vector. As a result, the first term in Eq.(\ref{Jy}) may be written in the form
\begin{eqnarray}\label{Jy1}
&&\mathrm{Tr}[j^y_{\mathbf{k}}G^{K1}_{\mathbf{k},\mathbf{k}}(\omega,\omega)]=\int dXdX^{\prime} \mathrm{Tr}\left[j^y_{\mathbf{k}}U(X)\times\right.\nonumber\\
&&\left.\tilde{G}^{K1}_{k_y}(X,X^{\prime})U^+(X^{\prime})\right]e^{i\omega(t-t^{\prime})-ik_x(x-x^{\prime})},
\end{eqnarray}
where $dX=dxdt$, $dX^{\prime}=dx^{\prime}dt^{\prime}$, and
\begin{equation}\label{GK1tilde}
\tilde{G}^{K1}_{k_y}(X,X^{\prime})=(\tilde{G}^{r}_{k_y}(X,X^{\prime})-\tilde{G}^{a}_{k_y}(X,X^{\prime}))f(\omega)\,.
\end{equation}
For brevity, in these equations the variable $X$  combines the coordinates $x$ and $t$. Since the system is uniform in the $y$-direction, Green's functions depend only on $y-y^{\prime}$ and, accordingly,  only on the single wave vector $k_y$.

Since the N\'{e}el vector varies in space and time much slower than Green functions, one can expand $\theta(X)$ and $\theta(X^{\prime})$ in the unitary matrices $U(X)=\exp(i\sigma_x\theta(X)/2)$ and $U^+(X^{\prime})=\exp(-i\sigma_x\theta(X^{\prime})/2)$ in Eq.(\ref{Jy1}) as
\begin{eqnarray}\label{theta}
&&\theta(X)=\theta(\bar{X})+\frac{1}{2}\partial_{\bar{t}}\theta(\bar{X})\tau+\frac{1}{2}\partial_{\bar{x}}\theta(\bar{X})\rho\nonumber\\
&&\theta(X^{\prime})=\theta(\bar{X})-\frac{1}{2}\partial_{\bar{t}}\theta(\bar{X})\tau-\frac{1}{2}\partial_{\bar{x}}\theta(\bar{X})\rho\,,
\end{eqnarray}
where $\bar{X}=(X+X^{\prime})/2$, $\rho=x-x^{\prime}$, and $\tau=t-t^{\prime}$. Further, by expanding  the SU(2) matrix up to linear terms with respect to gradients of $\theta$, and by taking into account that $\exp[i\sigma_x\theta(\bar{X})]$ commutes with $j^y_{\mathbf{k}}$,  Eq.(\ref{Jy1}) can be transformed to
\begin{eqnarray}\label{Jy12}
&&\mathrm{Tr}[j^y_{\mathbf{k}}G^{K1}_{\mathbf{k},\mathbf{k}}(\omega,\omega)]=\int dXdX^{\prime} \mathrm{Tr}\left[j^y_{\mathbf{k}}\tilde{G}^{K1}_{k_y}(X,X^{\prime})+\right.\nonumber\\
&&\left.j^y_{\mathbf{k}}\frac{i}{4}\{\tilde{G}^{K1}_{k_y}(X,X^{\prime}),\sigma_x\}(\partial_{\bar{t}}\theta(\bar{X})\tau+\partial_{\bar{x}}\theta(\bar{X})\rho)\right]\times\nonumber\\
&&e^{i\omega\tau-ik_x\rho}\,.
\end{eqnarray}
The first term in this expression corresponds to the adiabatic approximation with respect to the slowly varying N\'{e}el order. Within this approximation the Green functions vary fast as functions of their coordinate's difference $X-X'$, but their respective Fourier transforms still follow the relatively slow evolution of $\theta(\bar{X})$. As shown in Appendix A, this adiabatic term in Eq.(\ref{Jy12}) does not contribute in the current. Therefore, one should consider the second term, which contains gradients of $\theta(\bar{X})$. However, the gradient term, which is  proportional to $\partial_{\bar{x}}\theta(\bar{X})$, does not contribute to the electric current, at least up to linear expansions with respect to spatial gradients and the Rashba coupling constant (more details can be found in Appendix A).  At the same time, the contribution of $\partial_{\bar{t}}\theta(\bar{X})$ to the electric current is finite. In order to calculate it, first let us substitute Eq.(\ref{Jy12}) into Eq.(\ref{Jy}) and express $\tilde{G}^{K1}_{k_y}(X,X^{\prime})$ according  to Eq.(\ref{GK1tilde}). It is seen that the integration over $\omega$  involves the distribution function $f(\omega)$, which enters in the combination
\begin{equation}\label{domega}
\int d\omega e^{i\omega\tau}f(\omega)\tau=i\int d\omega e^{i\omega\tau}\frac{df}{d\omega}\,.
\end{equation}
At temperatures, which are  much lower than characteristic frequencies of the electronic system, one may set $df/d\omega=d\tanh(\omega/k_BT)/d\omega=2\delta(\omega)$ and the integral in Eq.(\ref{domega}) becomes simply $2i$. The importance of the step in the Fermi-Dirac statistical distribution signals that electron excitations near the Fermi level, which are produced by the moving DW, give rise to the electric current. This indicates the dissipative nature of this current.

The function $\tilde{G}^{K1}_{k_y}(X,X^{\prime})$ is expressed through unperturbed retarded and advanced functions  Eq.(\ref{G0}). They depend only on time and spatial coordinate differences, plus a term must be added to Eq.(\ref{G0}) due to the time dependent SOC $\mathbf{h}_{Rij}^{\overline{\nu}}(x,t)\bm{\sigma}=h_{Rij}^{y}(x,t)\sigma^y$ from  Eq.(\ref{Gtilde}). As shown in Appendix A, this term does not contribute to the current. Therefore, finally, only the unperturbed functions $\tilde{G}_{\mathbf{k}0}^{r(a)}(\omega)|_{\omega=0}$ should be taken into account. The integral over $X$ and $X^{\prime}$ in Eq.(\ref{Jy12}) can be transformed into the integral over $X-X^{\prime}$ and $\bar{X}$. Since the unperturbed Green's function depend only on the former variable, the integration over $\bar{X}$ involves solely $\partial_{\bar{t}}\theta(\bar{x},\bar{t})$. For a steady moving DW $\theta(\bar{x},\bar{t})$ depends on $\bar{x}$ and $\bar{t}$ in the form $\theta(\bar{x}-V\bar{t})$, where $V$ is the DW velocity. Therefore, $\partial_{\bar{t}}\theta(\bar{x},\bar{t})=-V\partial_{\bar{x}}\theta(\bar{x},\bar{t})$. Consequently, the integration of  $\partial_{\bar{x}}\theta(\bar{x},\bar{t})$ over $\bar{x}$ results in  $\theta(\infty)-\theta(-\infty)=\pm\pi$, where the plus sign corresponds to the N\'{e}el vector $n_z(-\infty)=1$. A more general situation with the arbitrary varying shape and velocity of a DW will be discussed in Sec.IV. By substituting in  Eq.(\ref{Jy12})  the Keldysh functions Eq.(\ref{GK1tilde}), where in the momentum representation $\tilde{G}^{r(a)}$ are given by Eq.(\ref{G0}),  we arrive at
\begin{equation}\label{Jy13}
J^{(1)}_y=-\frac{ieV}{8}\sum_{\mathbf{k}} \mathrm{Tr}[j^y_{\mathbf{k}}\{\tilde{G}^{r}_{\mathbf{k}0}(\omega)-\tilde{G}^{a}_{\mathbf{k}0}(\omega),\sigma_x\}|_{\omega=0}] \,.
\end{equation}

Next, let us consider the current $J^{(2)}_y$, which originates from the function $G^{K2}$. The current is given by  the second term in Eq.(\ref{Jy}), where $G^{K2}$ is represented by Eq.(\ref{GK2}). In the latter equation the integrand involves the function  $V_{\mathbf{Q}}(\omega^{\prime},\omega^{\prime}-\Omega^{\prime})$ which, in turn, is proportional to the difference of distribution functions $f(\omega^{\prime})-f(\omega^{\prime}-\Omega^{\prime})$, as it can be seen from Eq.(\ref{V}). In this equation the frequency $\omega^{\prime}$ is determined by  fast electron dynamics, in contrast to  $\Omega^{\prime}$ which is given by a relatively slow rotation of spins during DW motion. Therefore,  $f(\omega^{\prime})-f(\omega^{\prime}-\Omega^{\prime})\simeq \Omega^{\prime} df/d\omega^{\prime}  =2\delta(\omega^{\prime})\Omega^{\prime}$. Similar to Eq.(\ref{Jy13}), the delta-function fixes the the electron energy at the Fermi level due to the step-like energy dependence of the electron distribution function. Also, the function $V_{\mathbf{Q}}(\omega^{\prime},\omega^{\prime}-\Omega^{\prime})$ in Eq.(\ref{V}) becomes proportional to $\Omega^{\prime} \mathbf{n}_{\mathbf{Q}}(\Omega^{\prime})\bm{\sigma}$, that in the space-time representation can be written as $\partial_t \mathbf{n}(x,t)\bm{\sigma}$. Hence, the function $V_{\mathbf{Q}}(\omega^{\prime},\omega^{\prime}-\Omega^{\prime})$ contains the small nonadiabatic gradient term which should be taken into account only within the linear approximation. This simplifies the further analysis, because  the functions $G^r$ and $G^a$ in Eq.(\ref{GK2}) may be calculated within the adiabatic approximation by neglecting the gradient terms in Eq.(\ref{Gtilde}). Moreover, the space-time dependent SOC $\mathbf{h}_R^{\overline{\bm{\nu}}}(x,t)\bm{\sigma}$ in Eq.(\ref{Gtilde}) also does not contribute to the second term of Eq.(\ref{Jy}) (see Appendix A). Hence, in Eq.(\ref{GK2}) the retarded and advanced functions may be expressed through unperturbed functions Eq.(\ref{G0}), as $G^{r(a)}=U\tilde{G}_0^{r(a)}U^+$. Finally, as shown in Appendix A, we arrive at
\begin{equation}\label{Jy2a}
J^{(2)}_y=-\frac{eVJS}{2}\sum_{\mathbf{k}} \mathrm{Tr}[j^y_{\mathbf{k}}\tilde{G}^{r}_{\mathbf{k}0}(0)\sigma_y\tau_z\tilde{G}^{a}_{\mathbf{k}0}(0)] \,.
\end{equation}
This equation, together with Eq.(\ref{Jy13}) are the key results of the present study. The calculation of the total current $J_y=J^{(1)}_y+J^{(2)}_y$ is presented in detail in Appendix A. Note, that the sign of the current in Eqs.(\ref{Jy13}) and (\ref{Jy2a}) depends on the sign of the N\'{e}el order  at $x=-\infty$, providing that the positive direction of the $x$-axis coincides with the direction of the DW velocity. In this Section it was assumed that $n_z(-\infty)=1$. The sign of Eqs.(\ref{Jy13}) and (\ref{Jy2a}) depends also on DW's chirality. As was noted above, the latter is determined by the sign of the exponent in the unitary matrix $U=\exp(\pm i\sigma_x\theta(x,t)/2)$. Since the time derivative of this exponent enters in Eqs.(\ref{Jy13}) and (\ref{Jy2a}), the "$\pm$" chirality signs will also appear  in these equations. For a Bloch DW these signs correspond to the axial angles of the  N\`{e}el vector $\phi=\pm\pi/2$ . This result can be presented in a different form. It is easy to see that the $x$-component of the vector $(\mathbf{n}\times\partial_t\mathbf{n})_x=-\sin\phi\partial_t\theta$. Hence, the latter expression changes its sign when the chirality changes with the replacement of $\phi=\pi/2$ by $\phi=-\pi/2$. Since Eqs.(\ref{Jy13}) and (\ref{Jy2a}) are determined by the integral of $\theta$ over a space-time region which depends on the length of  electric contacts, it is convenient to introduce for these equations  the prefactor
\begin{equation}\label{C}
C=\frac{1}{\pi T}\int_0^Ldx\int_0^Tdt (\mathbf{n}\times \partial_t\mathbf{n}\times \mathbf{e}^z)_y
\end{equation}
instead of the DW velocity $V$. Indeed, for a steady moving DW it is easy to see from this equation that $C=V \sin\phi$, if $\theta|_{t=T}=\pi$.  Eqs.(\ref{Jy13}) and (\ref{Jy2a}) were calculated for one particular case of $\phi=\pi/2$. By taking the $x$ projection of the vector product in Eq.(\ref{C}) this equation can be also applied to the N\'{e}el DW.  This equation is valid for arbitrary moving domain walls, not only for steady moving ones. Moreover, the total charge which is pumped by DW during the time interval $T$ is a topological invariant which does not depend on how this wall moved through the device. We shall return to discussion of this  topic in Sec. IV.

As shown in Appendix B the approximate analytic result may be obtained in the most interesting case when the chemical potential is close to the van Hove singularity, so that $\mu \sim \Delta \ll t$ and  the Fermi line is close to the square. Within the linear approximation with respect to the Rashba SOC the current $J_y$ is given by
\begin{equation}\label{J1a}
J_y=-e\frac{2V}{a\pi}\frac{ \lambda_Rr^2}{\mu}\frac{1}{\sqrt{1-r^2}} \,,
\end{equation}
where $r=\Delta/\mu$. It should be noted that $\mu^2-\Delta^2$, which enters in the denominator of Eq.(\ref{J1a}), cannot be smaller than $\sim \mu\lambda_R$, because the perturbational expansion over $ \lambda_R$ is valid only at $\lambda_R \ll \mu-\Delta$.

\subsection{N\'{e}el DW}

In contrast to  the Bloch DW,  the N\'{e}el DW results in the current $j_x$ whose divergence $\partial_x j_x\neq 0$. This means that some charge could accumulate in the range of the DW. On the other hand, such a charge must be screened out by conduction electrons. This leads to the  electric potential $\varphi$ variation across the wall.  For a disconnected circuit this potential can be obtained from the simple equation
\begin{equation}\label{phi}
j_x-G \frac{d\varphi}{dx}=0\,,
\end{equation}
where $G$ is the metal conductivity. According to this equation, currents which are produced by the DW and the potential, compensate each other. By integrating this equation over $x$  we obtain
\begin{equation}\label{Jx}
J_x\equiv \int_{-L/2}^{L/2} dx j_x=G \Delta\varphi\,,
\end{equation}
where $\Delta\varphi$ is the potential difference between contacts. Since the distance between contacts $L$ is much larger  than the width of the DW, one may take  $L=\infty$ in Eq.(\ref{Jx}). This equation allows to express $\Delta\varphi$ in terms of $J_x$. Although this current can not be interpreted as a total current, like $J_y$, it is calculated in the same way as the latter. Namely, it is given by Eq.(\ref{Jy}), where the operator $j^y_{\mathbf{k}}$ should be substituted for $j^x_{\mathbf{k}}=\tau_x\partial_{k_x}(\epsilon_{\mathbf{k}}+\bm{\sigma}_y h^y_{R\mathbf{k}})$. The SU(2) rotation of  the Hamiltonian, which allows to align localized spins along the $z$-axis, becomes $U(x,t)=\exp(i\sigma_y\theta(x,t)/2)$. Accordingly, the spin-orbit interaction $\sigma_yh^y_R$ is invariant with respect to the unitary transformation, while $U(x,t)\sigma_xh^x_RU(x,t)^{+}$ varies in time and space. With these changes all calculations which are presented in Subsection IIIA can be applied to $J_x$, as well. As a result we arrive to $J_x$ given by a sum of Eqs.(\ref{Jy13}) and (\ref{Jy2a}), where $j^y_{\mathbf{k}}\rightarrow j^x_{\mathbf{k}}$. Also, $\sigma_x \rightarrow \sigma_y$ in Eq.(\ref{Jy13}), while $\sigma_y \rightarrow \sigma_x$ in Eq.(\ref{Jy2a}). In the unperturbed functions $G^{r(a)}_0$ the SOC term $\sigma_xh^x_R$ should be substituted for $\sigma_yh^y_R$. Finally, in both cases of Bloch and N\'{e}el domain walls Eq.(\ref{J1a}) is valid.  At the same time, the evaluation of the  voltage $\Delta\varphi$, which is induced by a N\'{e}el DW, involves the calculation of the conductivity $G$ within the same tight binding model  which was employed for calculations of $J_y$ in Sec.IIIA. The details of these calculations are presented in Appendix C. In order to evaluate $\Delta\varphi$, let use in expression Eq.(\ref{sigmafin}) for 2D conductivity  $G$ the following parameters: $t=5$ eV, $\Gamma=1$ meV, and $\sqrt{\mu^2-\Delta^2}/\mu=1/2$. We obtain $1/G\simeq 4\Omega$. As seen in Fig.2 the current $J\simeq 1$ nA. Hence, $\Delta \varphi \simeq $ 4nV.
\begin{figure}[tp]
\includegraphics[width=6.5 cm]{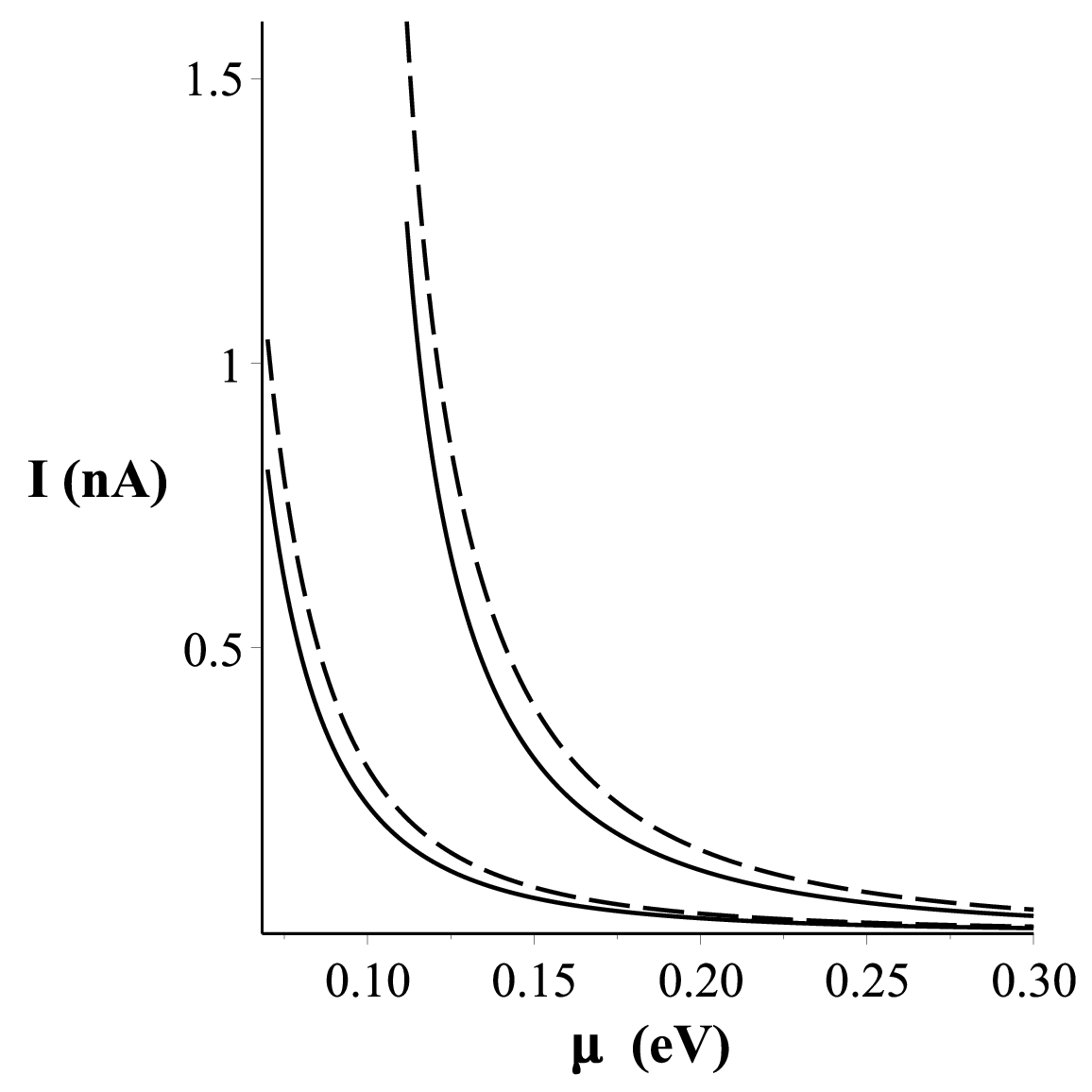}
\caption{The electric current $I$ which is induced by a moving Bloch domain wall, as a function of the chemical potential. The DW velocity $V=100$ m/s; the Rashba coupling constant $\alpha_R=5 $ meV; the lattice constant $a=0.5$ nm. Dashed curves are calculated according to approximate analytic formula Eq.(\ref{J1a}). Solid curves are numerically calculated, as a sum of Eqs.(\ref{J1fin}) and (\ref{J2fin}). The left and the right pairs of curves are calculated at $JS=0.05$ eV and $JS=0.1$ eV, respectively.} \label{fig2}
\end{figure}

\section{discussion}

The electric current produced by a Bloch DW, is shown in Fig.2 for two exchange interaction energies. The current increases when the chemical potential  approaches to the van Hove singularity (VHS) at $\mu=0$ and it  vanishes rapidly at larger values of $\mu$. As can be seen, the numerical results are close to those which follow from the approximate analytic expression    Eq.(\ref{J1a}). It is interesting that the analytic result does not depend on the  hopping parameter $t$. The current does not vary when the three parameters $\mu, J$, and $\alpha_R$ simultaneously change their scale, while remaining much less than $t$. One more interesting feature of the current is its independence on disorder, at least as long, as the elastic scattering rate $\Gamma \ll \Delta$ and $\Gamma \ll \alpha_R$. The current diverges at $\mu \rightarrow \Delta$, which is caused by the proximity of the chemical potential to the VHS. Since the perturbation expansion over $\alpha_R$ was employed, the cutoff $\sqrt{\mu^2-\Delta^2}>10\alpha_R$ was chosen in Fig.2 to remove this divergence. Though $\alpha_R$ in Fig.2 takes quite reasonable values, it would be interesting to extend the theory towards higher values of $\alpha_R\sim \Delta$. However, the space-time dependent part of SOC in Eq.(\ref{Gtilde}) poses a problem on this way. This coupling is a result of the unitary transformation of the initially uniform SOC. Therefore, its strength is determined by the same coupling constant $\alpha_R$, which enters into the space-time independent part of SOC. In the present study SOC was taken into account only as the linear perturbation. In this approximation the nonuniform part of SOC did not produce any effect, as shown in Appendix A. However, beyond this approximation the effect of the space-time dependent SOC is not easy to calculate. It requires a special study. Qualitatively, one might assume that the divergence of the current will saturate at $\alpha_R \sim \mu \sim SJ$.

The considered toy model is, in fact, a model for a two-dimensional antiferromagnet.  Though 2D antiferromagnets were studied experimentally, in particular, in van der Waals systems, there is no solid evidences that the stable fast moving domain walls exist there. On the other hand, such DW were observed in 3d AFM insulators. and ferrimagnets. Therefore, the considered model may be of use for a 2D metal which makes a contact to an AFMI. The main problem for its practical implementation is that it is necessary to have a commensurate interface between AFM and 2D metal lattices. Otherwise, the VHS enhancement of the current can not be reached. One might consider a monolayer of atoms which covalently bond to the magnetically compensated AFMI surface. With a proper choice of the adsorbate  this monolayer can form a metal layer, while the AFM order on the interface will stay intact. In other words,  the problem is to create the metallic surface state at the compensated surface of an AFM insulator.

At chosen in Fig.2 parameters,  the maximum current is in the range of 1nA. Also, the voltage which is induced by a N\'{e}el  DW is around 4nV, as evaluated in the end of Sec.IIIB. These numbers were calculated at the DW velocity 100 m/s which was experimentally observed in antiferromagnets \cite{Caretta,Siddiqui,Avci,Zhou,Velez,Gomonay2,Kim}. On the other hand, these values may be much higher for faster DW. For instance, DW velocities up to 4000 m/s  were observed in iron garnet \cite{Caretta} and 6000 m/s in compensated ferrimagnetic samples. \cite{Zhou} The effect might also be enhanced, as  mentioned above, at stronger SOC. One more possibility is to  fabricate a periodic array of lateral contacts, such that a moving DW could generate the periodic current. Such current might be enhanced by an electromagnetic resonator with the high Q-factor. For instance, if the distance between contacts in the array is 100 nm and the DW velocity is 100 m/s, the resonance frequency is 1GHz, which is just within the microwave region.

It is interesting to note that the pumped charge, or in other words, the time integrated current which is produced by a Bloch DW, when it passes through the racetrack, does not depend on how DW moves through the device. It can move with arbitrary acceleration. Indeed, since the pumped current is defined in  Eq.(\ref{Jy}) as an average over the time interval $T$, the total pumped charge is given by $JT$. As follows from Sec.IIIA, the current may be expressed as a sum of Eqs.(\ref{Jy13}) and (\ref{Jy2a}), where the velocity $V$ is substituted for the constant $C$, which is given by Eq.(\ref{C}). This constant is equal to $V$ for a steady moving DW, but it could be calculated in case of its arbitrary motion, as well. However,  the prefactor $1/T$ in Eq.(\ref{C}) is not uniquely defined in this case. Luckily, it vanishes in the expression for the pumped charge $JT$, where the factor $CT$ enters. It is easy to see that $CT$ is a topological invariant which does not depend on how the polar angle $\theta(x,t)$ of the N\'{e}el vector $\mathbf{n}=(\sin\theta\cos\phi, \sin\theta\sin\phi,\cos\theta)$ depends on $x$ and $t$. From Eq.(\ref{C}) it follows that $CT/L$ is always equal to $\pm 1$, as long as the width of a DW in the space-time domain is much less than $L$ and $T$, respectively. One may define the charge quantum $\mathcal{Q}$ by writing the pumped charge in the form $JT=(CT/L)\mathcal{Q}$. Therefore, several domain walls pump the charge $(N^+ - N^-)\mathcal{Q}$, where $N^+$ and $N^-$ are the number of DW with positive and negative chiralities, respectively.

The electric current pumping by a moving DW is, in a certain sense, reverse to the to spin orbit torque effect which can be produced by the electric current \cite{Zelezny,Gomonay2}. This torque may be caused by the staggered spin density which is accumulated in AFM due to the SOC assisted conversion of the electric current into the spin polarization. In the considered bilayer system, due to the interface exchange interaction,  the electric current of spin-orbit coupled electrons of the 2D metal gives rise to the staggered spin density which, in turn, diffuses into the  AFMI and produces there the torque effect on a DW. However, the AFMI film must be very thin for the interface diffusion to have an effect on DW within the whole 3D AFMI sample. In contrast, the considered here pumping effect, which is produced by a DW, does not  depend critically on the AFMI film thickness. Note, that a high DW speed due to the spin orbit torque effect may be reached at rather high current densities in the 2D film. For comparison, high DW velocities were observed at current densities more than 1A/m$^2$ in a bulk metal, adjacent to a compensated ferrimagnet \cite{Siddiqui}.


\appendix

\section{Calculation of the current produced by a domain wall}

The electric current which is produced by a Bloch DW is represented by two terms in Eq.(\ref{Jy}). Accordingly, these two currents will be denoted as $J^{(1)}$ and  $J^{(2)}$. The similar expression for  the N\'{e}el wall is obtained by the  replacement $y\rightarrow x$ in Eq.(\ref{Jy}). In this section we calculate the current and analyse all possible effects which originate from gauge fields and the time dependent Rashba field.

\subsection{Calculation of $J^{(1)}$}

The integrand of Eq.(\ref{Jy}), which corresponds to $J^{(1)}$, is given by Eq.(\ref{Jy12}).  Let us consider the contribution in the current of the first term in Eq.(\ref{Jy12}). It is easy to see that by substituting this term in Eq.(\ref{Jy}) we obtain the same equation with the Green function $G^{K1}$ being substituted for $\tilde{G}^{K1}=(\tilde{G}^{r}-\tilde{G}^{a})f(\omega)$. So modified Eq.(\ref{Jy}) results in zero current when SOC is absent. At the relatively weak SOC, such that $\alpha_R\ll \mu$, $\tilde{G}^{K1}$ may be expanded up to linear terms with respect to the Rashba field $\mathbf{h}^{\nu}_{Rij}$ and $\mathbf{h}^{\overline{\bm{\nu}}}_{Rij}(x_0,t)$ in Eq.(\ref{Gtilde}).
The corresponding corrections will be denoted as $\delta_{h}\tilde{G}^{r(a)}_{\mathbf{k},\mathbf{k}}(\omega,\omega)$ and $\delta_{\bar{h}}\tilde{G}^{r(a)}_{\mathbf{k},\mathbf{k}}(\omega,\omega)$, respectively. The former correction (its Fourier expansion over spatial coordinate and time differences) is given by the second term in Eq.(\ref{G0}), while the latter one can be similarly expressed as
\begin{eqnarray}\label{deltaGhbar}
&&\delta_{\bar{h}}\tilde{G}^{r(a)}_{\mathbf{k},\mathbf{k}}(\omega,\omega)=\int_0^Tdt\int_{-\frac{L}{2}}^{\frac{L}{2}}dx g^{r(a)}_{\mathbf{k}}(\omega)\times\nonumber\\
&&(\mathbf{h}^{\bar{\nu}}_{R\mathbf{k}}(t,x)\bm{\bar{\nu}})(\bm{\sigma}\bm{\bar{\nu}})\tau_xg^{r(a)}_{\mathbf{k}}(\omega)\,,
\end{eqnarray}
where $L$ is the length of the sample and $g_{\mathbf{k}}^{r(a)}(\omega)$ is given by Eq.(\ref{G0A}). The SOC corrections $\delta_{h}\tilde{G}^{r(a)}_{\mathbf{k},\mathbf{k}}(\omega,\omega)$ and $\delta_{\bar{h}}\tilde{G}^{r(a)}_{\mathbf{k},\mathbf{k}}(\omega,\omega)$ should be substituted in Eq.(\ref{Jy}) and their trace with the current operator $\nabla_{\mathbf{k}} \epsilon(\mathbf{k})\tau_x$ must be taken (the spin dependent current should not be taken into account in the linear approximation with respect to SOC). It is evident that such a trace is zero, by taking into account the spin dependence of $g^{r(a)}$ in Eq.(\ref{G0A}).

Further, we extend the perturbation expansion to include the  interaction of electrons with the gauge fields in  Eq.(\ref{Gtilde}).  The first order SOC correction stems from the interaction $\{\mathcal{A}_x,v^{xy}\}$. It is proportional to $\alpha_R$, due to the anomalous spin dependent velocity $v^{xy}$. Therefore, it might contribute  in the first perturbational order with respect to the spin-orbit interaction. This correction is similar to Eq.(\ref{deltaGhbar}), where $(\mathbf{h}^{\bar{\nu}}_{R\mathbf{k}}(t,x)\bm{\bar{\nu}})(\bm{\sigma}\bm{\bar{\nu}})\tau_x$ should be substituted for $\{\mathcal{A}_x,v^{xy}\}$.  It is easy to see that this term does not contribute to Eq.(\ref{Jy}) in the case of the Bloch DW, because in this case $\mathcal{A}_x \sim \sigma_x$, while $v^{xy}=\nabla_{k_x}h^y \sigma_y$, so that their anticommutator is zero. However, for the N\'{e}el  wall the corresponding anticommutator is finite, because $\mathcal{A}_x \sim \sigma_y$. Nevertheless, due to symmetry the result becomes zero after integration over wavenumbers in Eq.(\ref{Jy}). The reason is that the current $\nabla_{\mathbf{k}}\epsilon_{\mathbf{k}}$ changes sign at $\mathbf{k}\rightarrow -\mathbf{k}$, while $v^{xy}$ and $g_{\mathbf{k}}$ stay invariant. As a result, the integration of the odd function gives zero. One more first-order term originates from  the gauge interaction $\mathcal{A}_xv^x$, while the spin-orbit interaction is not taken into account in the Green function. Instead, this interaction contributes to the current operator $j^i_{\mathbf{k}}$. However, such a SOC dependent current $\partial_{k^i}\mathbf{h}_{\mathbf{k}}\bm{\sigma}$ is the even function of $\mathbf{k}$, while $\mathcal{A}_xv^x$ is the odd one. Therefore such a correction should not be taken into account.

Second-order perturbations involve products of the  gauge interactions $\mathcal{A}_0$ and $\mathcal{A}_xv^x$ in Eq.(\ref{Gtilde}) with the SOC operators  $\mathbf{h}^{\nu}_{R\mathbf{k}}\bm{\sigma}\tau_x$, or $\mathbf{h}^{\bar{\nu}}_{R\mathbf{k}}\bm{\sigma}\tau_x$. Let us first consider terms which include $v^x\mathcal{A}_x$. The latter is an odd function with respect to the inversion $\mathbf{k} \rightarrow - \mathbf{k}$, as well as  the Rashba field $\mathbf{h}_{R\mathbf{k}}$. Since the current $\nabla_{k_x}\epsilon_{\mathbf{k}}$ is also the odd function, the integration over $\mathbf{k}$ results in zero. At last, the interaction with the field $\mathcal{A}_0$ does not depend on spin. Therefore, its product with the SOC interaction will result in the zero trace. By summing all terms of  the perturbation expansion one can conclude that  the first term in Eq.(\ref{Jy12}) does not contribute to the current. Note, that this result is valid for both, Bloch and N\'{e}el  domain walls.

Now we consider the second term in Eq.(\ref{Jy12}). This expression is proportional to time and spatial gradients of the N\'{e}el order. Since we take into account only leading terms  with respect to such gradients, it is sufficient to consider only corrections to Green functions which are associated with SOC. Let us  first consider the term which is proportional to the time derivative of $\theta(x,t)$. It was shown in Sec.IIIA that for a degenerate electron gas this term results in a $\delta(\omega)$ frequency dependence of Eq.(\ref{Jy12}) and leads to Eq.(\ref{Jy13}) for the current $J^{(1)}$. This current is associated with the time independent SOC ($h^x_{R\mathbf{k}}\sigma_x$ for the Bloch DW). The linear in SOC terms enter into Eq.(\ref{Jy13}) through the spin dependent contribution $\nabla_{k_y}h^x \sigma_x\tau_x$ to the current operator  $j^y_{\mathbf{k}}$ and through the Rashba term in the functions $\tilde{G}^{r(a)}_{\mathbf{k}0}$, which are given by the second term in Eq.(\ref{G0}). By taking into account that $[\sigma_x\tau_x, g^{r(a)}_{\mathbf{k}}]=0$ and that the spin independent part of the current $j^y_{\mathbf{k}}$  is given by $\tau_x\nabla_{k_y}\epsilon_{\mathbf{k}}$,  Eq.(\ref{Jy13}) can be transformed to
\begin{eqnarray}\label{J1A}
&&J^{(1)}_y=-\frac{ieV}{8}\sum_{\mathbf{k}} \mathrm{Tr}[\nabla_{k_y}h^x_{R\mathbf{k}} \sigma_x\tau_x\{g^{r}_{\mathbf{k}}(0)-g^{a}_{\mathbf{k}}(0),\sigma_x\}+\nonumber\\
&&\nabla_{k_y}\epsilon_{\mathbf{k}}\tau_x\{(g^{r2}_{\mathbf{k}}(0)-g^{a2}_{\mathbf{k}}(0))h^{x}_{R\mathbf{k}}\sigma_x\tau_x,\sigma_x\}] \,.
\end{eqnarray}
The relations $(1+\beta \hat{P}_{\mathbf{k}})^2=2(1+\beta \hat{P}_{\mathbf{k}})$, and  $(1+\beta \hat{P}_{\mathbf{k}})(1-\beta \hat{P}_{\mathbf{k}})=0$ are helpful for the calculation of the trace in this equation. In such a way Eq.(\ref{J1A}) may finally be  written as
\begin{eqnarray}\label{J1fin}
&&J^{(1)}_y=eV\pi\left[\frac{\partial}{\partial \mu}\sum_{\mathbf{k}}h^x_{R\mathbf{k}}(\nabla_{k_y}\epsilon_{\mathbf{k}})\delta(E_{\mathbf{k}}-\mu)-\right.\nonumber \\
&&\left.\frac{\sqrt{\mu^2-\Delta^2}}{\mu}\sum_{\mathbf{k}}(\nabla_{k_y}h^x_{R\mathbf{k}})\delta(E_{\mathbf{k}}-\mu)\right]\,.
\end{eqnarray}
Note, that although each term of this equation is finite at $JS=0$, their sum in this limit turns to zero, as it is easy to check by taking into account that at $J\rightarrow 0$ $\delta(E_{\mathbf{k}}-\mu)\rightarrow\delta(\epsilon_{\mathbf{k}}-\mu)$ and $\nabla_{k_y}\epsilon_{\mathbf{k}}\partial_{\mu}\delta(\epsilon_{\mathbf{k}}-\mu)=-\nabla_{k_y}\delta(\epsilon_{\mathbf{k}}-\mu)$.

Let us analyze the contribution of the time dependent SOC in the second term of Eq.(\ref{Jy12}), which is proportional to $\partial_t\theta(x,t)$. For this, one should take into account the SOC correction Eq.(\ref{deltaGhbar}) in the function $\tilde{G}^{K1}$.  For the Bloch DW the Rashba field $\mathbf{h}^{\overline{\bm{\nu}}}$ is proportional to $\mathbf{e}_y\sin\theta(t,x)+\mathbf{e}_z\cos\theta(t,x)$. Therefore, the corresponding Rashba interaction carries operators $\sigma_y\tau_x$ and $\sigma_z\tau_x$, while all other spin operators in Eqs.(\ref{Jy12}) and (\ref{deltaGhbar}) enter as $\sigma_x\tau_x$ and $\sigma_z\tau_z$. As a result the trace over spin and sublattice variables in Eq.(\ref{Jy12}) turns to zero. Therefore, the time dependent SOC does not contribute in Eq.(\ref{Jy12}).

Above results are related to the $\partial_t\theta$ term of  Eq.(\ref{Jy12}). Let us consider the effect of the spatial gradient $\partial_x\theta(x,t)$. The integration over $\rho=x-x^{\prime}$ in Eq.(\ref{Jy12}) produces the derivative $\nabla_{k_x}\tilde{G}^{K1}$. A further integration by parts  over wavenumbers in Eq.(\ref{Jy}) gives rise to a product of $\tilde{G}^{K1}$ and $\nabla_{k_x}j_{\mathbf{k}}^y$. The latter contains the spin independent even function of $\mathbf{k}$  and the spin dependent odd function which is proportional to SOC $\alpha_R$. At the same time, $\tilde{G}^{K1}$ is a sum of spin independent even function Eq.(\ref{G0A}) and spin dependent linear in $h_R$ odd function Eq.(\ref{deltaGhbar}), or the similar second term in Eq.(\ref{G0}). The terms in such a product which are linear in SOC are odd. Therefore, they do not contribute in the current. Thus, it may be concluded that $J^{(1)}$ is given by Eq.(\ref{J1fin}).

\subsection{Calculation of $J^{(2)}$}

The Green function $G^{K(2)}_{\mathbf{k.k}}(\omega,\omega)$ in Eq.(\ref{Jy}) is given by Eq.(\ref{GK2}) at $\mathbf{Q}=0$ and $\Omega=0$. Since the current $j^y_{\mathbf{k}}$ does not depend on $\omega$ this function may be integrated over $\omega$. It is given by
\begin{eqnarray}\label{GK2A}
&&\int\frac{d\omega}{2\pi}G^{K(2)}_{\mathbf{k.k}}(\omega,\omega)=\sum_{\mathbf{Q}^{\prime},\mathbf{Q}^{\prime\prime}}\int\frac{d\Omega^{\prime}d\omega^{\prime}d\omega}{8\pi^3}
G^r_{\mathbf{k,k}-\mathbf{Q}^{\prime}}(\omega,\omega^{\prime})\times\nonumber\\
&&V_{\mathbf{Q}^{\prime\prime}}(\omega^{\prime},\omega^{\prime}-\Omega^{\prime})
G^a_{\mathbf{k}-\mathbf{Q}^{\prime}-
\mathbf{Q}^{\prime\prime},\mathbf{k}}(\omega^{\prime}-\Omega^{\prime},\omega)\,
\end{eqnarray}
One may write this integral  in terms of SU(2) transformed Green functions. In the space-time representation it has the form
\begin{eqnarray}\label{GK2AX}
&&\int d\mathcal{V}U(X_1)\tilde{G}^{r}_{k_y}(X_1,X)U^+(X)V(X,X^{\prime})\times \nonumber\\
&&U(X^{\prime})\tilde{G}^a_{k_y}(X^{\prime},X_2)U^+(X_2)e^{-ik_x(x_1-x_2)}\,,
\end{eqnarray}
where $d\mathcal{V}=dXdX^{\prime}dX_1dX_2dt_1$, $X_1=(x_1,t_1)$, $X_2=(x_2,t_1)$, $X=(x,t)$ and $X^{\prime}=(x,t^{\prime})$. The function $V(X,X^{\prime})$ in Eq.(\ref{GK2AX}) can be written as
\begin{equation}\label{VA}
V(X,X^{\prime})=-SJ\tau_zf(t-t^{\prime})(\bm{\sigma}\mathbf{n}(t,x)-\bm{\sigma}\mathbf{n}(t^{\prime},x))\,,
\end{equation}
where $f(\tau)=\int d\omega f(\omega)\exp(-i\omega\tau)$. By denoting $t=\bar{t}+\tau/2$ and $t^{\prime}=\bar{t}-\tau/2$ one can expand the slowly varying in time function $\mathbf{n}(t,x)$, up to the linear with respect to $\tau$ term. Hence, $\mathbf{n}(t,x)-\mathbf{n}(t^{\prime},x)=\tau\partial_{\bar{t}}\mathbf{n}(\bar{t},x)$. In turn, the function $\partial_{\bar{t}}\bm{\sigma}\mathbf{n}(\bar{t},x)$ may be written in the  form
\begin{eqnarray}\label{derivative}
&&\partial_{\bar{t}}\bm{\sigma}\mathbf{n}(\bar{t},x)=\partial_{\bar{t}}\left(e^{i\sigma_x\theta(\bar{t},x)/2}\sigma_ze^{-i\sigma_x\theta(\bar{t},x)/2}\right)=\nonumber\\
&&e^{i\sigma_x\theta(\bar{t},x)/2}\partial_{\bar{t}}\theta(\bar{t},x)\sigma_ye^{-i\sigma_x\theta(\bar{t},x)/2}\,,
\end{eqnarray}
Since $V(X,X^{\prime})$ is proportional to the time gradient of $\theta$, in the leading approximation  one may neglect other possible gradient terms. Therefore, we have $U^+(X)\exp(i\sigma_x\theta(\bar{t},x)/2)\simeq 1$ and $\exp(-i\sigma_x\theta(\bar{t},x)/2)U(X^{\prime})\simeq 1$. Besides, one should take into account that   Eq.(\ref{GK2AX}) will be multiplied by the current operator $j_{k}$ in  Eq.(\ref{Jy}) and the trace of their product will be taken. Since $[j_{k},U(X_1)]=0$, the unitary matrices $U(X_1)$ and $U^+(X_2)$ can be combined under the trace operation as the product $U(X_1)U^+(X_2)\simeq1$. Further, let us apply the mixed Fourier transformation to Green functions in Eq.(\ref{GK2AX}), such that it is applied only to coordinate differences, but the explicit dependence on $(X_1+X)/2$ and $(X^{\prime}+X)/2$ is retained. Since the dependence on these "center-of-gravity" coordinates stems from slow variations of the spin texture of DW, it is possible to identify them with $x$ and $t$. Finally, by taking into account  Eqs.(\ref{VA}) and (\ref{derivative}), the expression Eq.(\ref{GK2A}) can be transformed to
\begin{eqnarray}\label{GK2Afin}
&&\int\frac{d\omega}{2\pi}G^{K(2)}_{\mathbf{k.k}}(\omega,\omega)= \frac{iSJ}{2\pi}\int d\omega dtdx
\tilde{G}^r_{\mathbf{k}}(\omega,t,x)\times\nonumber\\
&&\tau_z\sigma_y\partial_{t}\theta(t,x)\tilde{G}^a_{\mathbf{k}}(\omega, t,x)\frac{df(\omega)}{d\omega}\,.
\end{eqnarray}
At low temperatures $df(\omega)/d\omega \simeq 2\delta(\omega)$. Therefore, $\omega=0$ in $\tilde{G}^r$ and $\tilde{G}^a$. In the linear with respect to SOC approximation  these functions are represented by $\tilde{G}^{r(a)}_{0\mathbf{k}}(\omega)+\delta_{\bar{h}}\tilde{G}^{r(a)}_{\mathbf{k}}(\omega, t,x)$, where the first and the second terms are given by Eq.(\ref{G0})  and Eq.(\ref{deltaGhbar}), respectively. For a Bloch DW in these equations $\mathbf{h}^{\nu}_{R\mathbf{k}}\bm{\sigma}=h^{x}_{R\mathbf{k}}\sigma_x$ and $\mathbf{h}^{\bar{\nu}}_{R\mathbf{k}}\bm{\sigma}=h^{y}_{R\mathbf{k}}(x,t)\sigma_y$. If only the former term is taken into account, the Green functions in Eq.(\ref{GK2Afin}) are space-time independent. Therefore, one may integrate $\partial_{t}\theta(t,x)$ over $x$ and $t$, as it was done in Sec.IIIA just above  Eq.(\ref{Jy13}). This integration results in  the factor $-VT\pi$ in Eq.(\ref{GK2Afin}). Further, by substituting Eq.(\ref{GK2Afin}) into Eq.(\ref{Jy}) we arrive at Eq.(\ref{Jy2a}) for the current $J^{(2)}$, if the time dependent SOC $h^{y}_{R\mathbf{k}}(x,t)\sigma_y$ is ignored. The current $j_{\mathbf{k}}$ in Eq.(\ref{Jy2a}) is a sum of  spin dependent and spin independent terms.  Since the former is proportional to $\alpha_R$ it is sufficient to combine this current with the unperturbed  functions $g^{r(a)}_{\mathbf{k}}(\omega)$ in Eq.(\ref{Jy2a}). At the same time, the conventional spin independent current should be taken together with the SOC dependent Green functions. By calculating the trace in  Eq.(\ref{Jy2a}) we obtain
\begin{equation}\label{J2fin}
J^{(2)}_y=-eV\pi\frac{\partial}{\partial \mu}\frac{\Delta^2}{\mu^2}\sum_{\mathbf{k}}h^x_{R\mathbf{k}}(\nabla_{k_y}\epsilon_{\mathbf{k}})\delta(E_{\mathbf{k}}-\mu)\,.
\end{equation}

In Eq.(\ref{J2fin}) the time dependent SOC is not taken into account. It is easy to see, however, that it does not contribute in $J^{(2)}$, at least in the linear with respect to $\alpha_R$ approximation. Indeed, by substituting Eq.(\ref{deltaGhbar}) in Eq.(\ref{GK2Afin}) instead of either $G^r$, or $G^a$, where $\bar{\nu}=y$ for a Bloch DW, after taking the trace of the obtained expression with the currant operator in Eq.(\ref{Jy}) one obtains the traces which look like  $\mathrm{Tr}[\nabla_{\mathbf{k}}\epsilon_{\mathbf{k}}\tau_x g^r_{\mathbf{k}}h^{y}_{R\mathbf{k}}(x,t)\sigma_y\tau_xg^r_{\mathbf{k}}\tau_z\sigma_yg^a_{\mathbf{k}}]$, plus permutations which involve $r,a$ superscripts, as well as $\sigma_x\tau_x$ and $\sigma_y\tau_z$ operators. According to Eq.(\ref{G0A}), the functions $g^{r(a)}$ are proportional to the projection operators $1\pm \hat{P}$. By employing the properties of these operators it is easy to see that such traces are zero.

Finally, we can conclude that the current $J_y^{(2)}$ is given by Eq.(\ref{J2fin}) and the total current is a sum of Eqs.(\ref{J2fin}) and (\ref{J1fin}).

\section{Analytic calculation of $J_y$}

Analytic results may be obtained at small $\mu \ll t$ when the shape of the  Fermi surface is close to the square. Let us first introduce dimensionless parameters. Accordingly, energies will be defined as $E/2t$ and lengths as $L/a$. Therefore, the dimensionless energy $\epsilon_{\mathbf{k}}=-(\cos k_x +\cos k_y)$, while the Rashba fields $h_x=-(\alpha_R/2t)\sin k_y$ and $h_y=(\alpha_R/2t)\sin k_x$. It is convenient to transform the integral over the Brillouin zone as
\begin{equation}\label{int}
\int \frac{dk_xdk_y}{4\pi^2}=\int \frac{dEdk_y}{4\pi^2}\frac{|E|}{|\sin k_x|\sqrt{E^2-\Delta^2}}\,,
\end{equation}
where $E=\pm\sqrt{(\cos k_x +\cos k_y)^2+\Delta^2}$ and $\Delta=JS/2t$.
Let us calculate first $J_y^{(1)}$. From  Eq.(\ref{J1fin}) this current can be expressed as
\begin{eqnarray}\label{J1B}
&&J^{(1)}_y=-\frac{eV\pi}{2}\frac{\alpha_R}{2t}\left[\frac{\partial}{\partial \mu}\int \frac{dEdk_y}{4\pi^2}\frac{|E|\sin^2 k_y\delta(\beta E-\mu)}{|\sin k_x|\sqrt{E^2-\Delta^2}}+\right.\nonumber \\
&&\left.\frac{\sqrt{\mu^2-\Delta^2}}{\mu}\int \frac{dEdk_y}{4\pi^2}\frac{|E|\cos k_y\delta(\beta E-\mu)}{|\sin k_x|\sqrt{E^2-\Delta^2}}\right]\,.
\end{eqnarray}
When the Fermi level lies in the lower energy band ($\mu < 0, \beta=-1$) we have $\cos k_x=\sqrt{\mu^2-\Delta^2}-\cos k_y$. It follows then that $\cos k_x \simeq -\cos k_y$, because $\mu \ll 1$ and $\Delta \ll 1$. Therefore, $\sin k_x \simeq \sin k_y$  when $0<k_x<\pi$ and $0<k_y<\pi$. Due to tetragonal symmetry it is sufficient to calculate the integrals just within this region and then multiply the result by four. By substituting $\sin k_x$ for $\sin k_y$ in the denominator of the first term in Eq.(\ref{J1B})  we obtain $\int_0^\pi dk_y \sin k_y=2$. In the second term of Eq.(\ref{J1B}) one should take into account the logarithmic singularities of $\sin^{-1} k_x$  at $k_x=0$ and $\pi$. By taking into account that at the Fermi line $\mu=-(\cos k_x +\cos k_y)$, these singularities can be regularized at $\mu<0$ by representing $\sin k_x$, as $\sin k_x=\sqrt{\sin^2 k_y+2|\mu|\cos k_y -\mu^2}$. However, $\cos k_y$ in the numerator has opposite signs at $k_x=0$ and $k_x=\pi$. Therefore, the singularities compensate each other, that results in integrals as small as $\mu$. A similar situation takes place when calculating the current $J_y^{(2)}$, which is given by Eq.(\ref{J1fin}). The same integrals over $k_y$ are present in $J_y^{(2)}$ . By leaving in Eq.(\ref{J1fin}) and Eq.(\ref{J2fin}) only leading terms we arrive at Eq.(\ref{J1a}).

\section{Conductivity of a disordered antiferromagnet}

Let us consider the conductivity of a 2D metal with the proximity induced AFM order, when the chemical potential is close to the Van Hove singularity. The relatively weak spin-orbit coupling is ignored. It is assumed that the width of the domain wall is larger than the electron's mean free path, so that one may neglect gradients of the N\'{e}el order within DW. Therefore, in the absence of SOC the spin quantization axis can be chosen parallel to the local direction of the N\'{e}el vector. The disorder is represented by random uncorrelated shifts $u_i$ of the site energies. The corresponding pair correlation function is $\overline{u_iu_j}=U\delta_{ij}$. Within the Born approximation the retarded and advanced selfenergies are given by \cite{Rammer}
\begin{equation}\label{sigma}
\Sigma^{r(a)}_{\alpha\alpha}(\omega)=U^2\sum_{\mathbf{k}}g^{r(a)}_{\mathbf{k}\alpha\alpha}(\omega)\,,
\end{equation}
where the Green function $g^{r(a)}$ is given by Eq.(\ref{G0A}) and the subscript $\alpha$ is the sublattice index. At $\omega \ll \mu$ one may set $\omega=0$ in Eq.(\ref{sigma}). Then,  the  integration over $\mathbf{k}$  gives
 \begin{equation}\label{sigma2}
 \Sigma^{r(a)}=\mp i\pi U^2\frac{N_{\mu}}{\mu 2}\left(1+\tau_z\sigma_z\frac{JS}{\mu}\right)\,,
 \end{equation}
 where $N_{\mu}$ is the state density at the Fermi level. By adding  Eq.(\ref{sigma2}) to the Hamiltonian Eq.(\ref{Hk0}), where in the absence of SOC $\mathbf{h}_R=0$, the  equation for the averaged Green function $\bar{g}^{r}_{\mathbf{k}}(\omega)$ is obtained in the form
\begin{equation}\label{HSigma}
\left[\omega +\mu+i\Gamma -\epsilon_{\mathbf{k}}\tau_x-JS\tau_z\sigma_z\left(1-i\frac{\Gamma_0}{\mu}\right)\right]\bar{g}^{r}_{\mathbf{k}}(\omega)=1\,,
\end{equation}
where $\Gamma_0=\pi U^2N_{\mu}/2$. In the leading approximation with respect to $\Gamma_0/\mu$   the Green function is expressed from Eq.(\ref{HSigma}) as
\begin{eqnarray}\label{gC}
&&\bar{g}_{\mathbf{k}}^{r}=\frac{1}{2}\frac{1+\hat{P}_{\mathbf{k}}}{\omega+\mu-E_{\mathbf{k}}+i\Gamma_0(1+\Delta^2/\mu|\mu|)}+\nonumber\\
&&\frac{1}{2}\frac{1-\hat{P}_{\mathbf{k}}}{\omega+\mu+ E_{\mathbf{k}}+i\Gamma_0(1-\Delta^2/\mu|\mu|)}\,,
\end{eqnarray}
where $\hat{P}_{\mathbf{k}}=(\epsilon_{\mathbf{k}}\tau_x+JS\tau_z\sigma_z)/E_{\mathbf{k}}$. Note, that the imaginary energy $i\Gamma_0(1-\Delta^2/\mu|\mu|)$ in the denominator of the second term of  Eq.(\ref{gC}) is, in fact, equal to $i\Gamma_0(1+\Delta^2/\mu^2)$ , because at small $\omega$ and positively defined $E_{\mathbf{k}}$ this term  has a pole only at $\mu<0$.

The Kubo formula and Eq.(\ref{gC}) can be employed for the calculation of the conductivity. By this way we obtain
\begin{equation}\label{sigmax}
G=\frac{e^2}{2}\sum_{\mathbf{k}}\int \frac{d\omega}{2\pi} \mathrm{Tr}[v^x_{\mathbf{k}}\bar{g}_{\mathbf{k}}^{r}(\omega)v^x_{\mathbf{k}}\bar{g}_{\mathbf{k}}^{a}(\omega)]\frac{df(\omega)}{d\omega}\,,
\end{equation}
where $v^x=\tau_x\nabla_{k_x}\epsilon_{\mathbf{k}}$. For a degenerate Fermi gas $df(\omega)/d\omega=2\delta(\omega)$. By integrating Eq.(\ref{sigmax}) and taking the trace we arrive at
\begin{equation}\label{sigmax2}
G=\frac{e^2}{\pi}\sum_{\mathbf{k}}\frac{(\nabla_{k_x}\epsilon_{\mathbf{k}})^2(1-\frac{\Delta^2}{E^2_{\mathbf{k}}})}{(|\mu|-E_{\mathbf{k}})^2+\Gamma^2}\,.
\end{equation}
where $\Gamma=\Gamma_0(1+\Delta^2/\mu^2)$. Near the van Hove singularity the integration over $\mathbf{k}$ involves the same type of integrals, as in the previous section, because $(\nabla_{k_x}\epsilon_{\mathbf{k}})^2=\sin^2k_x$. By employing  the same approximations, as in Appendix B the conductivity is obtained in the form
\begin{equation}\label{sigmafin}
G=\frac{4te^2}{|\mu|\pi^2}\frac{\sqrt{\mu^2-\Delta^2}}{\Gamma}\,.
\end{equation}
\end{document}